# Evidence for a Spectroscopic Sequence Among SNe Ia

Peter Nugent[1], Mark Phillips[2], E. Baron[1], David Branch[1] and Peter Hauschildt[3]

astro-ph/9510004    2 Oct 1995

[1]Dept. of Physics and Astronomy, University of Oklahoma, 440 W. Brooks, Rm 131, Norman, OK 73019-0225; nugent,baron,branch@phyast.nhn.uoknor.edu

[2]Cerro Tololo Inter-American Observatory, National Optical Astronomy Observatories, Casilla 603, La Serena, Chile; mphillips@noao.edu. CTIO/NOAO is operated by the Association of Universities for Research in Astronomy, Inc., (AURA) under cooperative agreement with the National Science Foundation

[3]Dept. of Physics and Astronomy, Arizona State University, Tempe, AZ 85287-1504; yeti@sara.la.asu.edu

## ABSTRACT

In this *Letter* we present evidence for a spectral sequence among Type Ia supernovae (SNe Ia). The sequence is based on the systematic variation of several features seen in the near-maximum light spectrum. This sequence is analogous to the recently noted photometric sequence among SNe Ia which shows a relationship between the peak brightness of a SN Ia and the shape of its light curve. In addition to the observational evidence we present a partial theoretical explanation for the sequence. This has been achieved by producing a series of non-LTE synthetic spectra in which only the effective temperature is varied. The synthetic sequence nicely reproduces most of the differences seen in the observed one and presumably corresponds to the amount of $^{56}$Ni produced in the explosion.

*Subject headings:* supernovae: general – supernovae: individual (SN 1991bg, SN 1986G, SN 1992A, SN 1989B, SN 1981B, SN 1994D, SN 1990N, SN 1991T)

– 3 –## 1. Introduction

Much of the recent work on SNe Ia has focused on the photometric differences between individual events. Correlations have been found between the absolute magnitudes at maximum light and the light curve decline rates. Relationships such as $\Delta m_{15}(B)$[4] vs. $M_B$ (Phillips 1993, Hamuy et al. 1995a) and the Light Curve Shape analysis (Riess, Press, & Kirshner 1995) highlight this work. A explanation for these photometric correlations has been provided by Höflich (1995) and Höflich and Khokhlov (1995) in which detailed hydrodynamic modeling has shown that SNe Ia which produce more $^{56}$Ni are brighter and have light curves which decay more slowly.

Spectroscopic correlations with absolute magnitude (or light curve decline rate) have also been searched for, but with mixed success to date. Pskovskii (1977, 1984) and Branch (1981) found evidence that the velocity of the minimum of the Si II $\lambda 6355$ absorption near maximum light was correlated with decline rate, but modern data do not appear to support this claim (Wells et al. 1994). Wells et al. (1994) suggested that the velocity of the minimum of the Ca II H&K absorption at maximum light may be correlated with decline rate, while Fisher et al. (1995) showed that the visual absolute magnitude is correlated with the minimum ejection velocity of calcium as measured from the H&K lines in moderately late-time spectra. Several authors have also called attention to general trends in the relative strengths of certain lines (e.g., Si II and Ti II) in the spectra of SNe Ia suggesting that these are correlated with decline rate (Phillips 1993, Hamuy et al. 1995a, Branch, Fisher, & Nugent 1993). Nevertheless, a robust numerical analysis of these trends has not yet been performed. In this *Letter* we discuss the evidence for such a spectroscopic sequence, presenting specific relationships between $M_B$ and 1) the ratio of the strength of the Si II absorption features found near 5800 and 6150 Å, and 2) the flux difference across the Ca II H&K absorption blend. Through the use of detailed non-LTE calculations of the spectra of SNe Ia we provide partial explanations for these correlations. Additionally, we make note of several other trends seen in both the observed and synthetic spectral sequences.

---

[4]The parameter $\Delta m_{15}(B)$ is defined as the amount in magnitudes that the B light curve declines during the first 15 days following maximum light.

## 2. Observations

Figure 1 presents a spectroscopic series of SNe Ia near maximum light. The observational sequence is placed in order of decreasing luminosity (increasing $\Delta m_{15}(B)$) as found in Phillips (1993) with the flux scaled arbitrarily for easier viewing. (The synthetic spectral sequence will be discussed in the following section). By placing the spectra in this order several trends become readily apparent along the observed sequence:

1. A pronounced difference occurs with the Ti II feature near 4200 Å. This is the most distinctive feature in the blue part of the spectrum of SN 1991bg. One can also see this absorption feature in SN 1986G, and perhaps a hint of it in SN 1992A. It is not present in SNe 1981B, 1990N and 1991T. However, these SNe show a more subtle change at these wavelengths as the Fe II lines in SN 1981B give way to the Fe III features seen in SNe 1990N and 1991T.

2. Two features which decrease in strength as one moves up the sequence are the absorption troughs at 7500 and 8200 Å. These features are associated with O I, Mg II and Ca II.

3. The relative strengths of the two absorption features which form the distinctive "W" absorption feature at ∼5400 Å normally ascribed to S II are observed to vary in strength along the sequence. The red line is stronger than the blue in the faster-declining (lower-luminosity) SNe, but as one moves up the sequence the two

    lines become nearly equal in strength in SNe 1981B and 1990N, whereas the blue line finally grows stronger than the red in the spectrum of SN 1991T.

4. Another interesting trend is observed in the relative strengths of the apparent emission features located just to the blue and red of the 3800 Å Ca II H&K absorption trough. This ratio becomes increasingly stronger as one moves up the sequence until the "jump" across the absorption essentially disappears in SN 1991T.

5. The Si II absorption troughs at 5800 and 6150 Å are perhaps the most striking features to evolve in this sequence. Proceeding up the sequence, the strength of the absorption at 5800 Å decreases with respect to the one at 6150 Å.

Figure 2 illustrates the latter two correlations. The ratio of the two Si II absorption features was measured via the following procedure. First, the spectrum is put onto an $F_\lambda$ vs. wavelength scale and then de-redshifted to correct for the radial velocity of the host galaxy. Second, the narrow Na I D interstellar lines (if present) are removed from the spectrum. Third, three "continuum" points are chosen to perform the analysis of the relative strengths. These points are the maxima in the emission features found just blue of the 5800 Å trough, between the two Si II troughs and to the red of the 6150 Å trough. Line segments are drawn between adjacent continuum points and then the difference in flux between the corresponding line and the minimum for each trough is calculated. Figure 2 displays the absolute blue magnitude $M_B$ vs. the ratio of the fractional depth of the bluer absorption trough to the redder one (which we refer to as the Si II ratio, $\mathcal{R}$(Si II)). Table 1 lists these values and their references. Note that the absolute luminosities were derived using the Surface Brightness Fluctuations and IR Tully-Fisher distance moduli listed by Phillips (1993) and Richmond (1995) , and therefore reflect the "short" distance scale ($H_0 \simeq 85$ km s$^{-1}$ Mpc$^{-1}$). The errors quoted for each ratio are rough estimates based on uncertainties in the reddening (which produce an almost negligible difference) and the night to night evolution of the spectrum. For the latter we examined the change in the ratios over a 10 day period centered on maximum light. The errors do not include the effects of poorly calibrated spectra. One might be tempted to assume a linear correlation between $M_B$ and $\mathcal{R}(\text{Si II})$, but as will be shown in the next section the reason behind this relationship is rather complex and can not be easily broken down into a simple correlation.

The flux ratio across the Ca II H&K absorption trough was measured in a similar fashion. After placing the SN spectrum on an $F_\lambda$ vs. wavelength scale and de-redshifting to correct for the host galaxy radial velocity, the flux is measured at two wavelengths, the maxima of the emission features which straddle the Ca II absorption. The Ca II ratio, $\mathcal{R}(\text{Ca II})$, is then defined as the ratio of the flux of the redder feature to the flux of the bluer feature. Figure 2 shows this ratio plotted vs. $M_B$. As is clearly seen, the relationship is remarkably similar to that observed for the Si II ratio; indeed, a plot of $\mathcal{R}(\text{Ca II})$ vs. $\mathcal{R}(\text{Si II})$ reveals a high degree of correlation between the two measurements.

## 3. Theory

In addition to the observed spectra shown in Figure 1 are six synthetic spectra. These spectra were produced with the code `PHOENIX` 4.8 (Hauschildt 1992a, Hauschildt 1992b, Hauschildt 1993, Hauschildt et al. 1995a, Baron et al. 1995, Hauschildt & Baron 1995, Hauschildt et al. 1995b). This code accurately solves the fully relativistic radiation transport equation along with the non-LTE rate equations (for some ions) while ensuring radiative equilibrium (energy conservation). The following ions were treated in non-LTE: Na I (3 levels), Ca II (5 levels), Mg II (18 levels), O I (36 levels) and Fe II (617 levels). All of the spectra are based on the same model parameters (one which produced a reasonable fit to SN 1991bg). The models have the following parameters: $t_{\text{rise}} = 15$ days (number of days since explosion), $V_R = 7500$ km/s (the velocity at the reference radius) and $V_e = 900$ km/s (the e-folding velocity which defines the density profile by $\rho(r) = \rho_o e^{-V/V_e}$). The reference radius, $R$, is the radius where the continuum optical depth at 5000 Å is unity. The abundances are taken from model W7 (Nomoto, Thielemann, & Yokoi 1984), homogenized for $v \geq 8000$ km s$^{-1}$, except that the titanium abundance has been increased by a factor of 10. The synthetic sequence was generated by varying *only* the observed luminosity of the SN, $L$. For easy reference to the spectra, we define the effective temperature, $T_{\text{eff}}$, as follows: $T_{\text{eff}}^4 = L/(4\pi R_o^2 \sigma)$. This is an input parameter of our models but our analysis makes no assumption about the existence of a photosphere. The models are all in radiative equilibrium so of course the temperature structures differ between them.

This synthetic sequence ranges in $T_{\text{eff}}$ from 7400 to 11,000 K. With only this change the synthetic spectra are able to reproduce most of the differences seen in the observed spectra. It must be noted that a particular synthetic spectrum does not correlate one to one with an observed spectrum (better fits can be obtained by adjusting the parameters), rather the values for $T_{\text{eff}}$ were chosen such that the trends seen in the observed sequence would be highlighted in the synthetic spectra.

The synthetic sequence reproduces the Ti II feature and its vanishing evolution as one increases $T_{\text{eff}}$. The increase in $T_{\text{eff}}$ causes the ionization of Ti II and the loss of its strong optical signature. Also of note is the effect the temperature change has on Ca II. One can easily see the decrease in strength of the infrared triplet as one moves up the sequence. This is once again caused by ionization. While observed data are sparse in this region SNe 1991bg, 1981B and 1991T definitely show this effect. Also, the emission feature blue-ward of the 3800 Å Ca II H&K blend shows the same increase in flux as is found in the observed spectra. This is due both to the increase in overall flux in the blue and to the ionization of Fe II and Co II, which decreases the line blanketing in this region of the spectrum, as $T_{\text{eff}}$ is raised.

Even though Si II is treated in LTE, the synthetic sequence is able to reproduce a similar correlation as seen in the observed $M_B$ vs $\mathcal{R}$(Si II). At first thought this correlation makes little sense. The trough at 5800 Å is due to a 4P-5S transition while the one at 6150 Å comes from the 4S-4P transition. Since they share the 4P level, and the line at 5800 Å line is produced by a transition with a higher excitation energy, it would be natural to think that this line should increase in strength relative to the 6150 Å line as $T_{\text{eff}}$ is increased. The reason behind its decrease in strength as one moves up the sequence is due to a rather complex interaction of the lines with line blanketing from Fe II and Co II which increases the apparent strength of the 5800 Å line at lower $T_{\text{eff}}$ while at higher $T_{\text{eff}}$ Fe III and Co III line blanketing helps to wash out the feature. Note the nice correlation seen in the synthetic spectra for $\mathcal{R}$(Si II) and $\mathcal{R}$(Ca II) as the luminosity is increased (see Figure 1).

Turning our attention towards the photometry of the synthetic models we can further explore the similarities between the models and observations. Figure 3 contains a graph of the absolute magnitude, in various bands, vs. $T_{\text{eff}}$ for all of the models used to create the synthetic sequence. The most striking feature in this graph is the almost 5 magnitude change in $M_U$ from the coolest to hottest model. (Note that in a more realistic sequence, where the more luminous models would have a longer risetime and hence a larger radius; these differences would only be enhanced.) Since previous fits to SNe 1981B and 1992A have been good in the UV (Nugent et al. 1995a), it is not unreasonable to think that this is representative of what happens in the observed spectral sequence. Unfortunately, the number of quality observations in $U$ are rather limited. What we do have (Schaefer 1995) does indicate such a trend (Branch et al. 1995). One should also note that the range in absolute magnitude for all of the models decreases as one moves to redder photometric bands. This effect has manifested itself in Phillips (1993) where the slope of the correlation between $\Delta m_{15}(B)$ and $M_B$ is seen to be steeper than between $\Delta m_{15}(B)$ and $M_I$.

In addition Figure 3 displays the maximum-light colors of the synthetic models vs. $T_{\rm eff}$. This figure shows that the $U$ band is a powerful tool for discriminating the differences among SNe Ia. $U - B$ changes almost 2 magnitudes over this range of $T_{\rm eff}$ while the color differences seen in $V - R$ and $R - I$ barely span 0.5 magnitude. Also of note are the derived bolometric corrections for these models. [Note that we have corrected an error in our synthetic photometry that appeared in Baron et al. (1993) and Nugent et al. (1995a, 1995b) ; see Nugent et al. (1995c) ]. Models whose colors correspond to typical SNe Ia $[|B - V| < 0.2$ (Vaughan et al. 1995)], correspond to models with $T_{\rm eff} \gtrsim 9000$ K.

## 4. Discussion and Conclusions

We have shown that the observed photometric sequence of SNe Ia (Phillips 1993, Hamuy et al. 1995a, Riess, Press, & Kirshner 1995) manifests itself equally clearly as a spectral sequence. This fact may prove to be extremely useful to current attempts to measure the deceleration parameter, $q_o$, through the discovery and observation of z $\sim$ 0.3–0.5 SNe Ia. The problem facing these programs is to correct for the expected luminosity spread of the individual SNe. Until now, the only tool for doing this was to measure the initial decline rate of the light curve [One could perform a color cut and eliminate peculiar SNe Ia (Vaughan et al. 1995), but this could further deplete an already small, high z, sample of data]. However, if a spectrum is obtained at maximum light (which is required to verify that the SN is actually type Ia event), a measurement of either $\mathcal{R}$(Ca II) or $\mathcal{R}$(Si II) should, in principle, provide an independent estimate of the luminosity of the event. Of these two ratios, $\mathcal{R}$(Ca II) may prove to be the easier one to measure, particularly in the highest redshift SNe.

From our synthetic spectra, we conclude that the observed spectroscopic and photometric sequences of SNe Ia are due primarily to temperature differences. Since the supernova is almost entirely powered by the radioactive decay of $^{56}$Ni, the temperature differences are likely to be due to the total amount of $^{56}$Ni produced in the explosion. Undoubtedly abundance and velocity differences also play a role, but their effects are of secondary importance. We know, for example, that the titanium overabundance that we have used for the models in this paper is probably unique to underluminous SNe Ia such as SN 1991bg (Filippenko et al. 1992). Nevertheless, in our temperature sequence the effects of the titanium overabundance are masked by the ionization of Ti II as one moves to higher temperatures. Similarly, while our hot model bears a strong resemblance to the maximum light spectrum of SN 1991T, the features produced by intermediate mass elements are too strong, since these elements were underabundant in this supernova (Jeffery et al. 1992). Further progress in understanding this spectroscopic sequence more fully is likely to come from a more detailed study of the two extremes observed so far in the sequence, SN 1991bg and SN 1991T. This will be the subject of further work.

This work was supported in part by NASA grant NAGW-2999, a NASA LTSA grant to ASU, and by NSF grants AST-9115061 and AST-9417242. Some of the calculations in this paper were performed at the NERSC, supported by the U.S. DOE, and at the San Diego Supercomputer Center, supported by the NSF; we thank them for a generous allocation of computer time.

Table 1. Observed $\mathcal{R}$(Si II) and $\mathcal{R}$(Ca II).

| SN | $B$ | $\mu$ | $E_{B-V}$ | $M_B$ | $\mathcal{R}$(Si II) | $\mathcal{R}$(Ca II) | References |
|---|---|---|---|---|---|---|---|
| 1991T | 11.69(03) | 30.60(30) | 0.13(05) | -19.44(36) | 0.14(05) | 0.94(11) | 1,2,3 |
| 1981B | 12.03(05) | 30.50(30) | 0.10(05) | -18.88(37) | 0.16(05) | 1.42(11) | 4,2,3 |
| 1994D | 11.86(03) | 30.68(13) | 0.00(02) | -18.82(16) | 0.29(05) | 1.38(11) | 5-6,7,8 |
| 1990N | 12.73(03) | 31.40(30) | 0.00(02) | -18.67(31) | 0.16(05) | 1.14(11) | 9,2,3 |
| 1989B | 12.34(05) | 29.40(30) | 0.37(03) | -18.58(33) | 0.29(05) | 1.29(11) | 10,2,3 |
| 1992A | 12.57(03) | 30.65(11) | 0.00(02) | -18.08(14) | 0.38(05) | 1.58(11) | 11,2,3 |
| 1986G | 12.45(05) | 27.71(08) | 0.65(10) | -17.92(42) | 0.53(05) | 2.07(11) | 12,2,3 |
| 1991bg | 14.76(10) | 31.08(08) | 0.00(02) | -16.32(15) | 0.62(05) | 2.56(11) | 13,2,3 |

References. — (spectra, $\mu$, $B$ & $E_{B-V}$ ) (1) Phillips et al. 1992, (2) Phillips 1993, (3) Hamuy et al. 1995b, (4) Branch et al. 1983, (5) Meikle et al. 1995, (6) Patat et al. 1995, (7) Tonry 1995, (8) Richmond et al. 1995, (9) Mazzali et al. 1993, (10) Wells et al. 1994, (11) Kirshner et al. 1993, (12) Phillips et al. 1987, (13) Filippenko et al. 1992

---

This manuscript was prepared with the AAS LaTeX macros v4.0.

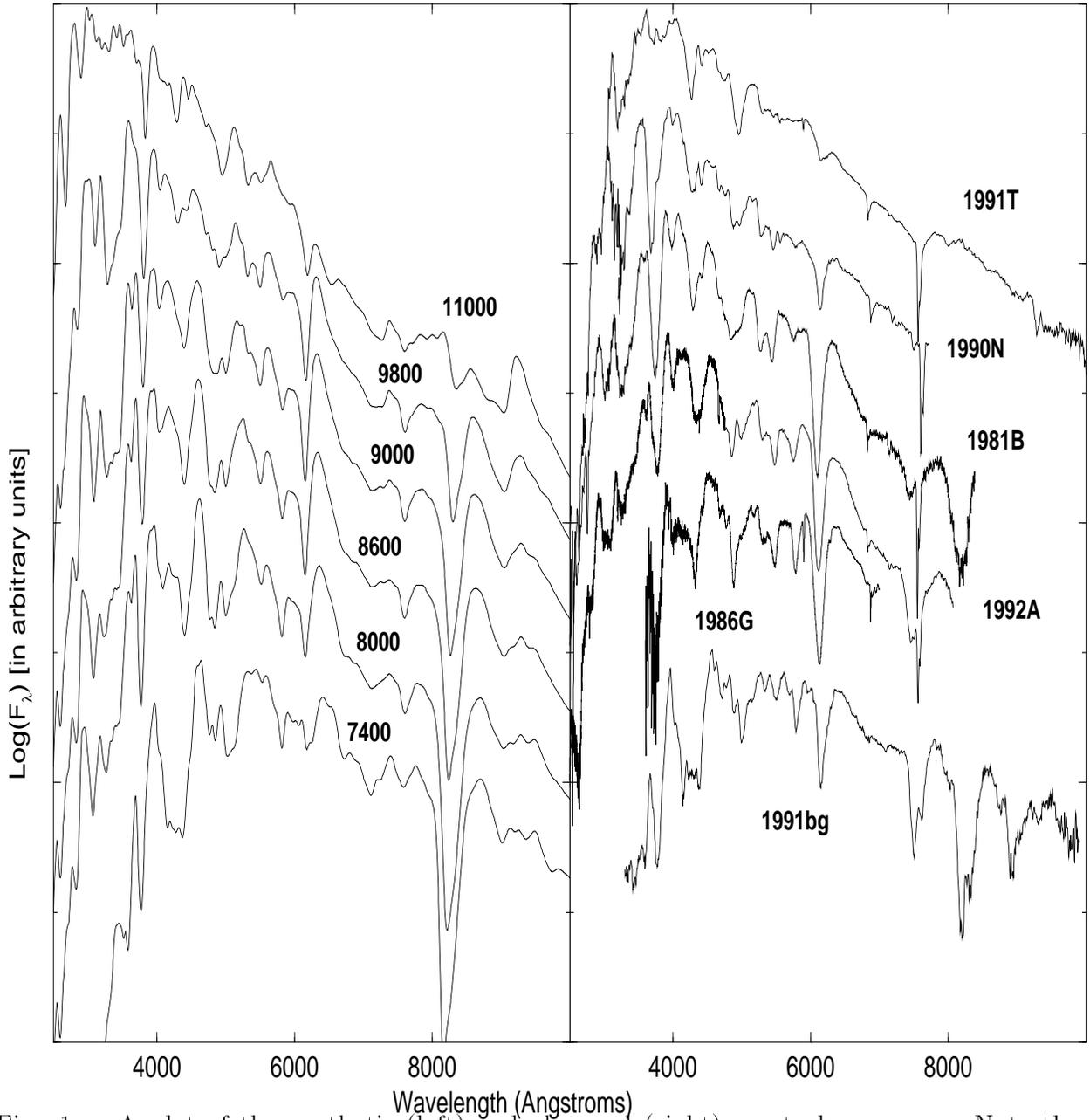

Fig. 1.— A plot of the synthetic (left) and observed (right) spectral sequences. Note the evolution in each graph of the following spectral features: Ca II at 3800 and 8200 Å, Ti II at 4200 Å and Si II at 5800 and 6150 Å.

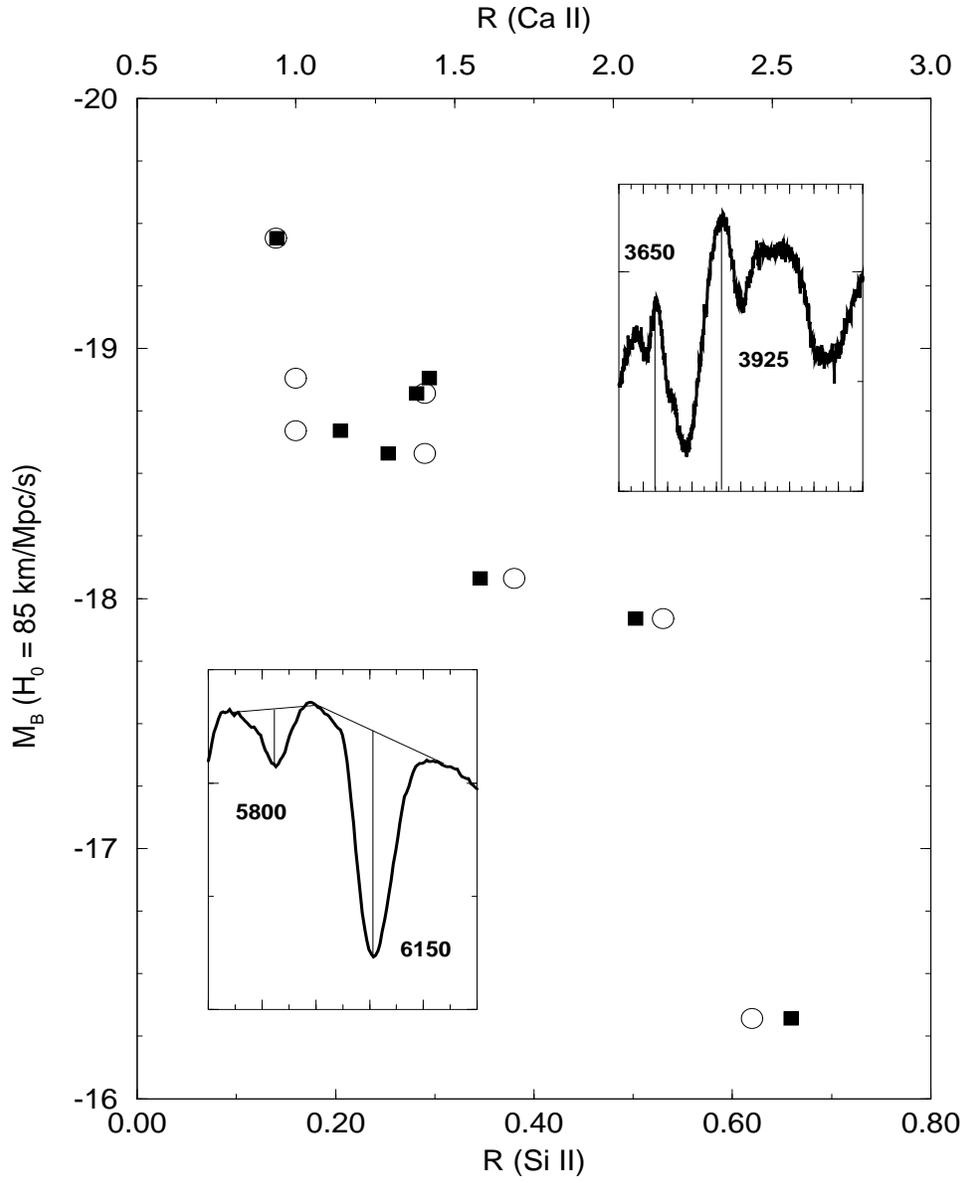

Fig. 2.— Observed $M_B$ vs. $\mathcal{R}$(Si II) (open circles) and $\mathcal{R}$(Ca II) (filled squares). The inset graphs illustrate how the ratios were measured. Error estimates can be found in Table 1

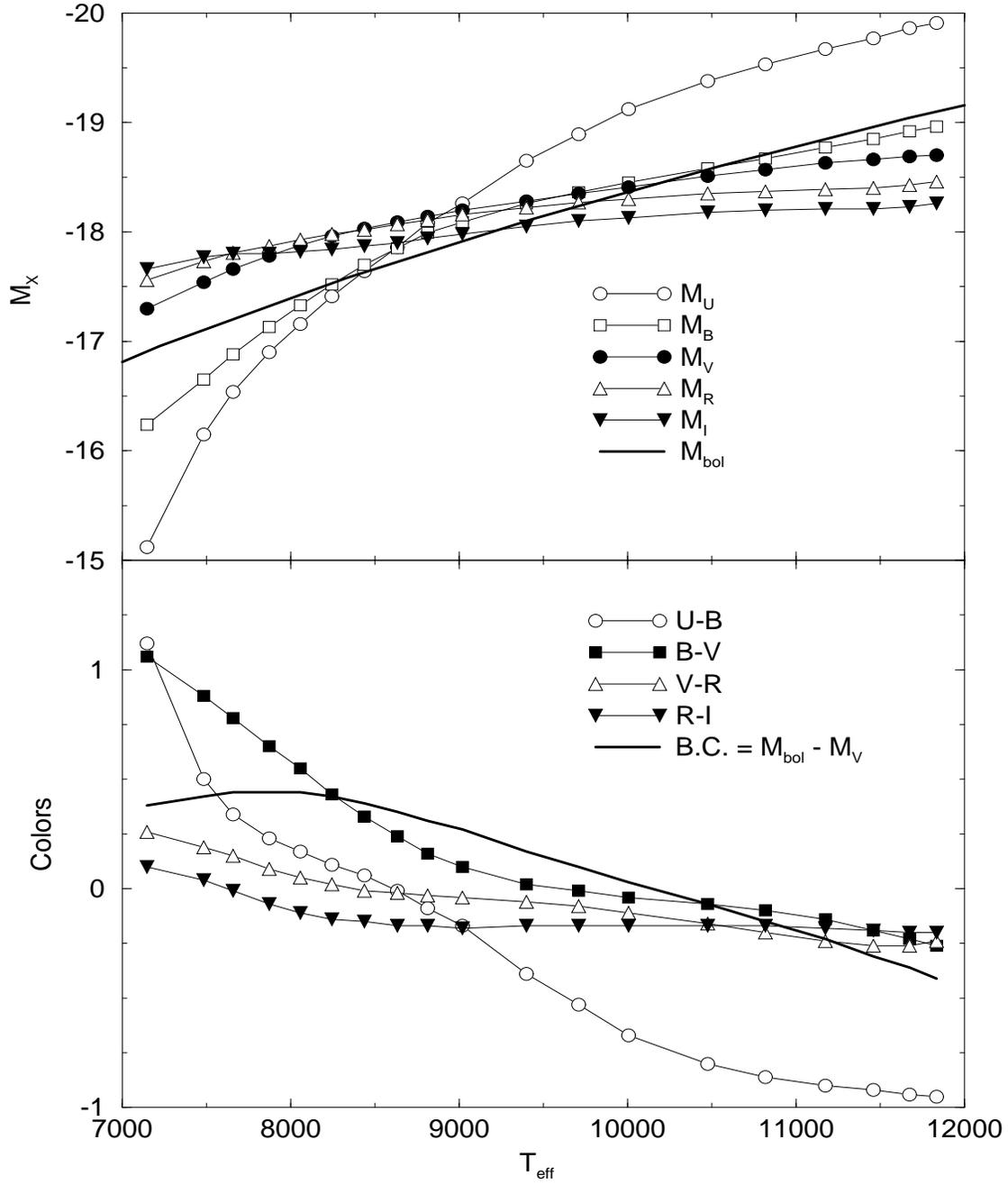

Fig. 3.— Photometry in $U$, $B$, $V$, $R$ and $I$ (top panel) and the colors (bottom panel) for the synthetic spectra.